# Finite Phase-separation FRET I: A quantitative model valid for bilayer nanodomains

*Frederick A. Heberle [a]\* and Gerald W. Feigenson[b]*

[a,b]Field of Biophysics, Cornell University, Ithaca, NY 14853

AUTHOR EMAIL ADDRESS: [a]fah2@cornell.edu, [b]gwf3@cornell.edu

*To whom correspondence should be addressed. Email: fah2@cornell.edu, Phone: (865) 576-8802.

**ABSTRACT**

Multicomponent lipid mixtures exhibit complex phase behavior, including coexistence of nanoscopic fluid phases in ternary mixtures mimicking the composition of the outer leaflet of mammalian plasma membrane. The physical mechanisms responsible for the small size of phase domains are unknown, due in part to the difficulty of determining the size and lifetime distributions of small, fleeting domains. Steady-state FRET provides information about the spatial distribution of lipid fluorophores in a membrane, and with an appropriate model can be used to determine the size of phase domains. Starting from a radial distribution function for a binary hard disk fluid, we develop a domain size-dependent model for stimulated acceptor emission. We compare the results of the model to two similar, recently published models.

**INTRODUCTION**

Förster resonance energy transfer (FRET) is an immensely useful tool for probing molecular length scales. The physical mechanism for FRET is a through-space interaction of electric oscillators in the near field: it occurs at distances greater than molecular contact, but much smaller than the wavelength of visible light. Resonance between electric dipoles in an excited-state donor molecule and ground-state acceptor molecule (which for the purposes of this work are fluorophores) results in simultaneous donor quenching and acceptor excitation, without emission of a donor photon. The accessibility of FRET to experimentalists is largely due to Förster, who correctly described the now well-known dependence of FRET efficiency on donor-acceptor separation distance, and who provided a valuable set of equations for quantifying FRET in many physically realistic scenarios (1). It is for this reason that the phenomenon, which was first observed well before Förster's contributions, today bears his name.

The inherent distance dependence enables FRET between freely diffusing membrane fluorophores to provide insight into the mixing behavior of lipids. This was recognized as early as 30 years ago, when three (nearly simultaneous) extensions to Förster's theory provided a quantitative description of FRET for a random, planar array of fluorophores (2-4). With the recent emergence of the lipid raft hypothesis, these models have been revisited in the context of coexisting membrane phases (5,6). Indeed, FRET is a particularly attractive tool for studying lateral organization in biological membranes because it can be employed in biologically relevant free-standing bilayers, in contrast to many other methods for probing small length scales that require a rigid bilayer support. The technology required to do FRET experiments is inexpensive and ubiquitous, and a wide variety of suitable probes are commercially available; consequently the literature for membrane FRET experiments is rapidly expanding.

As evidence mounts that the size of plasma membrane rafts may play an important biological role, the quantitative use of membrane FRET as a "molecular ruler" for phase domains has been explored (7,8). The exquisite sensitivity of FRET to precise details of molecular geometry demands precision in the analysis. Care must be taken to ensure that the geometric picture on which the model is based—that is, the spatial distribution of donors and acceptors—is an accurate description of the system. To this end, several groups have recently proposed modifications to the original models for membrane FRET to better account for energy transfer near domain boundaries, where the approximations of the early models are not valid (9,10). Such considerations are critically important for an accurate determination of domain size in the 2-50 nm range, a range that overlaps with many experimental and theoretical estimates of domain size in both cells and model membrane systems (11-13).

This paper provides a brief overview and comparison of the two most influential early models for membrane FRET, those of Wolber and Hudson (WH), and Fung and Stryer (FS) (3,4). An extension of the FS model to multibilayer systems is then proposed, and its application to the characterization of lamellarity in liposome preparations is discussed. We then address a crucial update to the model: a quantitative description of energy transfer efficiency for probes that partition between phase domains of arbitrary size. Included is a discussion of the recent work of Towles et al. and Brown et al. to extend the WH and FS models (respectively) to account for small domains, and the deficiencies in the latter model that motivated this work (9,10).

**METHODS**

*Monte Carlo simulations of transfer efficiency*

Refer to the companion paper for details of these simulations.

**RESULTS AND DISCUSSION**

This section is organized as follows:
1. We describe and compare two early models for membrane FRET, upon which nearly all subsequent work is based.
2. We propose and develop an extension to FS for the case of small domains: the Finite Phase-separation FRET (FP-FRET) model.

3. The FP-FRET model is compared to two existing models for FRET efficiency derived for the case of small domains.

*Comparison of two influential models for membrane FRET*

The theory of either Wolber and Hudson (WH) or Fung and Stryer (FS), published nearly simultaneously in the late 1970s, serves as the starting point for most subsequent treatments of membrane FRET (3,4). A survey of the literature reveals that the WH equations are by far the most commonly used, although a discussion of what differences exist between these models (if any) is difficult to find. The popularity of WH may be due to the greater generality of their presentation, including an easily computed approximation to the exact solution. In contrast, the FS model requires numerical integration, which was far more computationally expensive at the time that work was published than it is today. As the FS model serves as the basis for the theory in the rest of this paper, an explanation of this choice is in order.

By way of comparison, both derivations explicitly employed the following assumptions:

1. The concentration of excited-state donors is much less than the concentration of acceptors (the limit of moderate excitation intensity).
2. Donors and acceptors are randomly distributed in the membrane.
3. There is no change in the distribution of donor-acceptor distances (i.e., no lateral diffusion) on the timescale of fluorescence emission. In real membranes these timescales are separated by at least two orders of magnitude, lending validity to this assumption.
4. The Förster distance $R_0$, which quantifies the spatial extent of the FRET interaction for a particular donor-acceptor pair, is not a function of distance between a donor and acceptor.
5. There is no time-dependence to $R_0$ on the timescale of fluorescence decay (i.e., either the static or dynamic averaging limit holds). In practice, assumptions 4 and 5 allow all donor-acceptor interactions to be described by a single value of $R_0$.

In either model, the relative donor quenching due to FRET and the transfer efficiency are given by the following equations:

$$q_r = \int_0^\infty e^{-t/\tau_0} e^{-n_A S(t)} dt \qquad 1$$

$$E = 1 - q_r \qquad 2$$

where $n_A$ is the acceptor concentration (a surface density given in number of acceptors per unit area), and $S(t)$ is the so-called "energy transfer term". For FS, the energy transfer term has the form:

$$S_{FS}(t) = \int_{R_e}^\infty 2\pi r \left[1 - e^{-(t/\tau_0)(R_0/r)^6}\right] dr \qquad 3$$

while for WH, it is:

$$S_{WH}(t) = \pi R_0^2 (t/\tau_0)^{1/3} \gamma[2/3, (t/\tau_0)(R_0/R_e)^6] + \pi R_e^2 \left[1 - e^{-\left(\frac{t}{\tau_0}\right)\left(\frac{R_0}{R_e}\right)^6}\right] \qquad 4$$

where $\gamma$ is the incomplete gamma function given by:

$$\gamma(x, y) = \int_0^y z^{x-1} e^{-z} dz \qquad 5$$

Both models ultimately reveal dependence of donor quenching on three parameters: the surface density of acceptors $n_A$, the Förster distance $R_0$, and the distance of closest approach between a donor and acceptor $R_e$. Neither model has a closed form algebraic solution, so that the equations must be solved by numerical integration.

Given the identical assumptions of the two models and the similar forms of the equations, what then are the differences between the WH and FS models? First, the WH equations have a very simple power series expansion when the distance of closest approach between donor and acceptor is much less than $R_0$ (mathematically, when $R_e \sim 0$). The resulting infinite sum converges rapidly, making a precise computation of q$_r$ extremely fast and efficient. However, in real membrane systems, this approximation is rarely if ever valid: the closest-contact distance for donors and acceptors in the same leaflet is approximately the sum of the respective molecular radii (often not much smaller than $R_0$), and the closest-contact distance for acceptors in the opposing leaflet is often *greater* than $R_0$. An additional exponential term in Equation 1 is required to account for this contribution to donor quenching. In the more physically realistic geometries with nonzero $R_e$, both the WH and FS equations must be integrated numerically. As a second and more important difference, WH provided an approximate expression for donor quenching (with a stated accuracy of < 1%) in the form of a simple sum of two exponentials, and included a table of best-fit values for the pre-exponential and exponential fitting parameters over a range of acceptor concentrations. Although the double exponential form is orders of magnitude faster to compute than a numerical integral, either can now be calculated in a fraction of a second on any laptop computer.

*Finite Phase-separation FRET (FP-FRET): A model for FRET in the presence of small domains*

It is now clear that many lipid mixtures containing cholesterol, from the simple three-component model systems to the immensely complex plasma membrane, exhibit non-random mixing of lipid components over a wide range of size scales. In the case of unstimulated plasma membrane, while much evidence supports the presence of lateral domains (11-13), direct observation by conventional fluorescence microscopy has failed to detect large-scale phase separation. The thermodynamic nature of these lateral heterogeneities is the subject of much debate, and reliable measurements of size- and timescales of domains are critically important. FRET between freely-diffusing membrane probes is sensitive to domain sizes in the range of ~ 2-20 nm, given an appropriate model for the spatial arrangement of donors and acceptors.

Defining a one-size-fits-all geometry to model lateral domains in even "simple" three-component membranes is itself problematic, let alone for modeling rafts in the vastly more variable and chemically heterogeneous plasma membrane. A simple geometric model should be considered a first-order approximation for small phase domains. Nevertheless, the derivation outlined here can serve as a starting point for investigating more complicated geometries.

The following thermodynamic and geometric picture is used in the derivation:
1. Domains result from first-order phase separation of two liquids.

2. The donor and acceptor molecules are non-interacting, and distribute between the phases according to a well-defined partition coefficient.
3. At a given composition of interest, one of the liquid phases exists as circular domains dispersed in the other, surrounding liquid phase.
4. The circular phase domains are monodisperse with a radius $R$.
5. The domains are non-interacting: there are no special forces that cause domains to attract or repel, resulting in a random domain distribution.

It is worth stating explicitly the practical implications of these assumptions. By assumption 1, the total area fraction of domains can be determined mathematically from the lever rule and the average molecular areas of the pure phases, while assumption 2 allows us to treat the probes as being randomly distributed within the phase domains. Importantly, the model parameters resulting from assumptions 1 and 2 (phase fractions, molecular areas, and partition coefficients) can be determined by independent experiments. Assumptions 3 and 4 simplify the mathematics and minimize the number of fitting parameters (though it should be noted that a *distribution* of domain sizes can be accommodated with a fairly straightforward modification of the equations derived in this section). Assumption 5 ensures that the circular domains are randomly distributed within the surrounding phase. Together, assumptions 3-5 enable the introduction of an important concept in statistical mechanics—the pair correlation function—to model the spatial distribution of domains. As a final note, the derivation that follows is for energy transfer from donors to acceptors in the same plane. Using the results of the previous section and the assumption of cross-leaflet domain coupling, the equations in this section are easily modified to account for non-equivalent donor-acceptor planes and cross-leaflet energy transfer. This situation is explicitly addressed in the companion paper.

We first note that the FS (or WH) equations are easily modified for the "infinite phase separation" condition—that is, for the case where $R$ (domain radius) approaches infinity. (In practice, the "infinite phase separation" case is indistinguishable from a domain diameter greater than ~ 40 times the Förster distance, see the companion paper). Here, the idea is to simply treat the coexisting phase domains in the sample as if they could be separated by some mechanical means, and their quenching measured separately. One only needs to know the fraction of total donor and acceptor found in each phase; FRET is then independently calculated for each donor pool using the FS equations, and summed:

$$E = f_D^d E^d + f_D^s E^s \qquad 6$$

where $E^d$ and $E^s$ are the phase-specific transfer efficiencies calculated from Equations 1-3. The fraction of total donors in each phase is calculated by mass balance, from the mole fraction of domain phase $\chi^d$ (determined from the lever rule) and the partition coefficient of the donor $K_D^d$:

$$f_D^d = K_D^d \chi^d / (1 - \chi^d + \chi^d K_D^d) \qquad 7$$

$$f_D^s = (1 - \chi^d) / (1 - \chi^d + \chi^d K_D^d) \qquad 8$$

where $K_D^d \in (0, \infty)$ is defined so that values greater than unity indicate preference for the domain phase. It should be noted that the absolute donor concentration does not enter into these equations: transfer efficiency is independent of absolute donor concentration in the limit of low excited state donor concentration (i.e., excited state donors are not competing for acceptors). Additionally, the phase-specific acceptor surface density must be used in Equation 1, and is similarly calculated by mass balance:

$$n_A^d = K_A^d \chi_A / [a^d (1 - \chi^d + \chi^d K_A^d)] \qquad 9$$

$$n_A^s = \chi_A / [a^s (1 - \chi^d + \chi^d K_A^d)] \qquad 10$$

$$n_A^\infty = \chi_A / [a^d \chi^d + a^s (1 - \chi^d)] \qquad 11$$

where $n_A^d$ and $n_A^s$ are the respective acceptor surface densities in the domain and surround phases, $a^d$ and $a^s$ are the respective molecular areas of the domain and surround phases, $\chi_A$ is the total acceptor mole fraction, and $K_A^d$ is the acceptor partition coefficient with values >1 indicating preference for the domain phase. The bulk acceptor surface density (Equation 11) is the area-weighted average acceptor surface density, and can be thought of as the acceptor density of a thin shell at an infinite distance from any particular donor in the bilayer. It is included here for completeness.

Why do the FS or WH models fail when domains are small? After all, we have explicitly assumed that the properties of the two coexisting phases are no different from the case of "macroscopic" phase separation: the lever rule still holds, and probes still partition between phases with a well-defined $K_P$. To answer this question, we must first recognize that the population of donors located within domains consists of two sub-populations (to a first-order approximation), each with different average environments: donors that are *near* domain boundaries "see" an acceptor environment that is different from donors located well inside the domain. What is meant by "near" a boundary? The inverse sixth-power distance dependence of FRET ensures that less than 1% of energy transfer occurs to an acceptor located further than 2.2 $R_0$ from a given donor (for the largest common $R_0$, this is about 18 nm). By this criterion, any donor located within ~20 nm of a domain boundary has a non-negligible contribution to quenching from acceptors across the phase boundary (that is, in the coexisting phase). This is in fact true even in the limit of infinite phase separation: the difference is that for micron-sized domains, these problematic donors—those located within 20 nm or so of a domain boundary—make up a negligible fraction of the total donor pool. As domains get smaller and domain perimeter increases, the problematic donors become the majority, and so we must find a way to quantify their local acceptor environment.

The key to a solution is a subtle point implicit in Equations 9-10: the ensemble-averaged acceptor concentration in the neighborhood of a donor located *far from a phase boundary* does not depend on distance from the donor. This is not the case for donors located *near a phase boundary*. If we can mathematically account for the distance-dependence in local acceptor concentration (in other words, determine the correct *r*-dependence for Equations 9-10, given the assumptions of the model), we can simply insert those functions into the existing FS machinery:

$$S(t) = \int_{R_e}^{\infty} \langle n_A(r) \rangle 2\pi r \left[1 - e^{-(t/\tau_0)(R_0/r)^6}\right] dr \qquad 12$$

The remaining derivation focuses on arriving at the ensemble-averaged acceptor density functions $\langle n_A^d(r) \rangle$ and $\langle n_A^s(r) \rangle$. We imagine observing a donor located within a domain over time, occasionally pausing to measure the distances to all acceptors in its vicinity. Over long periods of observation, a normalized histogram of these distance measurements will converge on $\langle n_A^d(r) \rangle$, the ensemble-averaged acceptor surface density function for donors *inside* domains. If instead we choose to observe a donor located *outside* a domain (in the *surround* phase), we will arrive at the function $\langle n_A^s(r) \rangle$. We will show that through mass balance, the latter function can be obtained from the former.

We approach the problem by first ignoring the probe molecules, and considering only the domains: from a given reference point within a domain, what is the probability that some randomly chosen point at a distance $r$ will also be located inside a domain? The problem, stated in this way, closely resembles a classical problem in statistical mechanics with a very large literature, the two-dimensional fluid. The simplest 2D fluid—monodisperse, non-interacting particles, also known as the "hard-disk fluid"—is characterized by a single parameter, the disk packing fraction $f$. For our purposes the disk packing fraction is equivalent to the domain area fraction, defined as:

$$f = a^d \chi^d / [a^d \chi^d + a^s (1 - \chi^d)] \qquad 13$$

Figure 1 shows Monte Carlo snapshots of a hard disk fluid for two packing fractions, $f = 0.2$ (panel *A*) and $f = 0.5$ (panel *B*). Of central importance to theoretical treatments of the hard disk fluid is the pair correlation function $g(r; f)$, also referred to as the radial distribution function or RDF (14). The RDF for unit-radius disks with packing fraction $f = 0.5$ is shown in Figure 2*A*. The RDF describes the relative probability of finding another disk center some distance $r$ away from the center of a reference disk. Characteristics of the RDF for a hard-disk fluid are:

1. It is zero between 0 and $2R$, a consequence of the hard-core repulsion (disks cannot overlap).
2. It is normalized to the number density of disks in the system, so that at large distances g($r$) approaches 1.
3. There are oscillations in probability (about 1) at short distances that increase in magnitude and decay length as the disk packing fraction increases. The peaks occur approximately at integer multiples of the disk diameter, and reflect "shells" of nearest-neighbor disks surrounding the reference disk centered at 0.

It is interesting to note that the short-range oscillations in $g(r)$ are an *apparent* attraction that is purely statistical in nature, as there is no explicit interaction between disks except a hard-core repulsion. They are the consequence of local hexagonal order that arises at higher number densities, as disks are packed ever more tightly into the system.

As it turns out, there is no analytical solution for $g(r)$ in even-numbered dimensions (15): the function must be obtained either by Monte Carlo simulation (as was done for the curve in Figure 2A, see the companion paper for simulation details) or through an approximate expression. With this in mind, we begin with the RDF as given, for a particular packing fraction $f$ and scaled distance $r' = r/R$, where $R$ is the domain radius. We will use the function notation $g(r'; f)$ where the independent variable (here, $r'$) is separated from any fixed parameters by a semicolon. This notation serves as a reminder that $g(r)$, or any function derived from it, is specific to some packing fraction $f$. The number density for unit-radius domains is $f/\pi$, and the local number density $\gamma(r'; f)$ is defined as:

$$\gamma(r'; f) = (f/\pi) g(r'; f) \qquad 14$$

$\gamma(r')$ for $f = 0.5$ is shown in Figure 2B. $\gamma(r')$ is useful because it gives the expected number of domain centers $N$ found in an annulus of width $dr'$, located at a distance $r'$ from the center of the reference domain:

$$\langle N \rangle_{r'} = 2\pi r' dr' \, \gamma(r'; f) \qquad 15$$

We now return to the question that prompted our diversion into statistical mechanics: What is the probability that a randomly chosen point at a distance $r'$ from the reference domain center will *itself* be located inside a domain? This probability is equal to the *domain surface coverage* at $r'$, which we will call $\sigma^d$, and to answer the question we must know how many domains we expect to observe at $r'$. Equation 15 tells us the number of *domain centers* we will see at $r'$, but these are not the only domains contributing to the surface coverage at $r'$: in fact, any unit-radius domains located between $r'-1$ and $r'+1$ will contribute to the average at $r'$. Figure 3 shows how a unit-radius domain near $r'$ will contribute to surface coverage of a thin shell of radius $r'$ centered at the origin, through an angle $2\theta$. The infinitesimal domain surface coverage at $r'$ contributed by a domain centered at $r'+x$ is proportional to the area of the annular segment of width $dr'$ between $-\theta$ and $+\theta$, and to the expected number of domain centers at $r'+x$ (given by Equation 15):

$$d\sigma^d \propto 2r'\theta(r',x)dr'\, 2\pi(r'+x)\gamma(r'+x;f)dx \qquad 16$$

$$\theta(r,x) = \cos^{-1}\left[\frac{2r^2 + 2rx + x^2 - 1}{2r(x+r)}\right] \qquad 17$$

Integrating Equation 16 for all contributing values of *x*, and normalizing to the total area of the annulus located at $r'$ (that is, $2\pi r' dr'$) gives:

$$\sigma^d(r';f) = \begin{cases} 1, & 0 \le r' < 1 \\ 2\int_{-1}^{1} (r'+x)\,\gamma(r'+x;f)\,\theta(r',x)dx, & r' \ge 1 \end{cases} \qquad 18$$

Equation 18 gives the probability of finding a domain at a distance $r'$ from the *center* of some reference domain. The equation is shown for packing fraction $f = 0.5$ in Figure 2C. A more useful function gives the probability of finding a domain at a distance $r'$ from a *randomly chosen point* within the reference domain. This is the ensemble-averaged domain surface coverage $\langle \sigma^d(r';f) \rangle$, valid for all donors inside domains in the long-time average. It is found by averaging Equation 18 over all positions within the reference domain:

$$\langle \sigma^d(r'; f) \rangle = 1/\pi \int_0^1 \rho \, d\rho \int_0^{2\pi} \sigma^d(\sqrt{r^2 + \rho^2 + 2r\rho \cos \varphi}; f) d\varphi \qquad 19$$

Equation 19 is shown for packing fraction $f = 0.5$ in Figure 2D. We can now express the acceptor density near an average donor located inside a domain as:

$$\langle n_A^d(r; f, R) \rangle = n_A^d \langle \sigma^d(Rr'; f) \rangle + n_A^s(1 - \langle \sigma^d(Rr'; f) \rangle) \qquad 20$$

where we have reintroduced the absolute distance $r = Rr'$. This is the first of the two sought-after equations, the distance-dependent counterpart of Equation 9. The second is obtained by mass balance: rearranging Equation 9 and replacing $n_A^d$ with Equation 20, we arrive at an expression for a distance-dependent "apparent" partition coefficient of the acceptor into the domain phase:

$$K_A^{d,app}(r; f, R) = \frac{a^d(1 - \chi^d)}{\chi_A/\langle n_A^d(r; f, R) \rangle - \chi^d a^d} \qquad 21$$

Inserting this expression into Equation 10 gives the distance-dependent counterpart to Equation 20, valid for donors located in the surround phase:

$$\langle n_A^s(r; f, R) \rangle = \chi_A / [a^s(1 - \chi^d + \chi^d K_A^{d,app}(r; f, R))] \qquad 22$$

Equations 20 and 22 give the dependence of acceptor surface density on distance from an average donor in the two donor pools: those located in domains, and those located in the surround. They are inserted directly into the integrand of the FS energy transfer integral (Equation 3) to complete the small domain model, taking the place of what had previously been a constant local acceptor density (no dependence on *r*).

Figure 4A shows Equation 20 (red curve) and Equation 22 (blue curve) for $f = 0.4$, and for (arbitrarily chosen) values of acceptor partition coefficient ($K_A^d = 10$), mole fraction ($\chi_A = 0.002$), and phase molecular areas ($a^d = 0.675$, $a^s = 0.45$). Several interesting features are evident in these curves. At the shortest distances, the "local vicinity" of a given donor must be either inside a domain for $\langle n_A^d(r) \rangle$, or in the surround for $\langle n_A^s(r) \rangle$: consequently the limiting values of these curves at $r = 0$ are simply the domain or surround acceptor densities defined by Equations 9-10. At the other extreme $r = \infty$, the acceptor density is equal to the bulk density defined by Equation 11: regardless of where a particular donor is located (in a domain or in the surround), at long distances the acceptor environment must look like the bulk average of the system. At intermediate distances the curves oscillate about the bulk acceptor density, which reflects the locally hexagonal ordering of randomly packed domains. Fig 4*B* shows a histogram of the acceptor surface density for the domain (red circles) and surround (blue circles) phases, determined by Monte Carlo simulations (see the companion paper for simulation details). These simulations demonstrate the validity of Equations 20 and 22.

*Comparison to existing models*

Similar FRET models have recently been published elsewhere (9,10). In the model of Towles and Dan (TD), the WH equations were modified with a directly simulated domain surface coverage function similar to Equation 19, and the resulting model was evaluated with simulated data (9). An advantage of the present work is that the domain surface coverage function is derived through mathematical manipulation of the pair correlation function for hard disks $g(r)$, a more general approach that allows for the use of existing functional forms for $g(r)$. The theoretical literature for two-dimensional fluids is extensive, and many well-characterized approximations to $g(r)$ exist (15). In principle, any functional form for $g(r)$ can be used, including those derived for polydisperse (16) or interacting disks (17). Indeed, a long-range interaction between fluid phase domains has been suggested as a potential mechanism for stabilizing small domains (18): using the approach outlined above, the effect of such an interaction on the spatial distribution of probes can be accounted for via $g(r)$.

A second model, published by the same research group, was derived from the FS equations. In contrast to the TD model (which was only tested against simulated data), the Brown model was used to extract domain size for experimentally obtained FRET data in several model bilayer mixtures (DMPC/cholesterol, DPPC/DOPC/cholesterol, and DPPC/POPC/cholesterol) (10,19). For this reason, it deserves special scrutiny. In addition to the largely geometric assumptions found in the TD and FP-FRET models, several additional (and critically important) simplifications were made:

1. For donors located inside domains, the quenching contribution to acceptors within the same domain is neglected. Effectively, this limits analysis to experimental systems where donors and acceptors partition into opposite phases, and where acceptor partition is exceptionally strong.
2. Donors inside domains were assumed to reside on a ring at 2/3 $R$ (the mass-weighted "average" radial position of an ensemble of randomly distributed donors).
3. The acceptor surface density was assumed to be constant (and equal to its surround-phase density, viz. Equation 18) throughout the membrane.

The result is a highly specific model that is valid only in a small and poorly defined subset of the overall parameter space: when the FRET probes partition into different phases, and acceptor partition is strong, and domain area fraction is very small (the infinitely dilute domain limit, realized only near tieline endpoints).

Figure 5 compares predictions of the FP-FRET (solid line), TD (dashed line), and Brown (dotted line) models for two sets of parameters chosen to mimic experimental data on an Ld + Lo tieline (Table 1). Figure 5*A* shows FRET efficiency (E) for strong donor and acceptor partition into the Ld phase, while Figure 5*B* shows the case of strong donor preference for Lo and strong acceptor preference for Ld. Also shown in both plots are data generated from Monte Carlo simulations using the same parameter sets (circles, see the companion paper for simulation details). The predictions of the FP-FRET and TD models are similar for the length of the tieline, with small differences in E ($< \sim 1\%$) seen at some compositions. The differences may be due to differences in the domain surface coverage functions. In both cases, surface coverage data were simulated at discrete values of $f$ and $r$, and an integrable function suitable for use in the model was generated from the data (see the companion paper). The number of data points in both the $f$ and $r$ dimensions is considerably higher in the present work, which should produce more accurate results. Even so, the observed differences in the predictions of the two models are small relative to the accuracy of the domain size analysis reported in the companion paper.

The limitations of the Brown model are evident in Figure 5. As Brown et al. note, their model is valid only for opposite probe partitioning, and only in the regime of small domain packing fractions (i.e., small domain area fractions): these restrictions correspond to regions near the tieline endpoints in Figure 5*B*. Despite this significant caveat, the model was used to estimate domain sizes over the entire Ld + Lo coexistence region of two ternary mixtures. Figure 5 demonstrates that E predicted from the Brown model deviates significantly from simulated values along the length of the tieline, suggesting that the region of validity is perhaps more narrow than the authors appreciated.

**CONCLUSIONS**

We have derived a model for transfer efficiency between freely diffusing membrane probes in phase-separated bilayers that is valid in the regime of finite-sized phase domains. The FP-FRET model yields predictions that are similar to those of a recently published model derived under similar assumptions, and which closely agree with FRET efficiency data generated with Monte Carlo simulations over the entire range of domain packing fractions. The small differences between these models are unlikely to result in substantial differences in recovered domain size. In contrast, a third model (Brown) was shown to deviate significantly from the simulated data, due to a highly restrictive set of assumptions. Consequently, domain sizes recovered from experimental data in the Ld + Lo coexistence regions of DPPC/DOPC/chol and DPPC/POPC/chol are likely to contain significant errors. These results graphically illustrate the importance of geometric considerations in models for transfer efficiency in membranes.

**Table 1** Parameters used in energy transfer simulations.

| Parameter | Description | REE | RRE |
|---|---|---|---|
| $a_{Ld}$ | Average molecular area of Ld phase (nm$^2$) | 0.675 | |
| $a_{Lo}$ | Average molecular area of Lo phase (nm$^2$) | 0.45 | |
| $\chi_{perc}$ | Phase percolation threshold | 0.5 | |
| $\chi_A$ | Mole fraction acceptor | 0.001 | |
| $\tau_0$ | Donor fluorescence lifetime (ns) | 1.0 | |
| $\varphi$ | Relative donor quantum yield | 1.0 | |
| $r$ | donor, acceptor radius (nm) | 0.5 | |
| $d$ | donor, acceptor transverse location (nm) | 1.3 | |
| $R_0$ | Förster distance (nm) | 5.5 | |
| $K_D$ | donor partition coefficient | 0.1 | 10 |
| $K_A$ | acceptor partition coefficient | 0.1 | 0.1 |

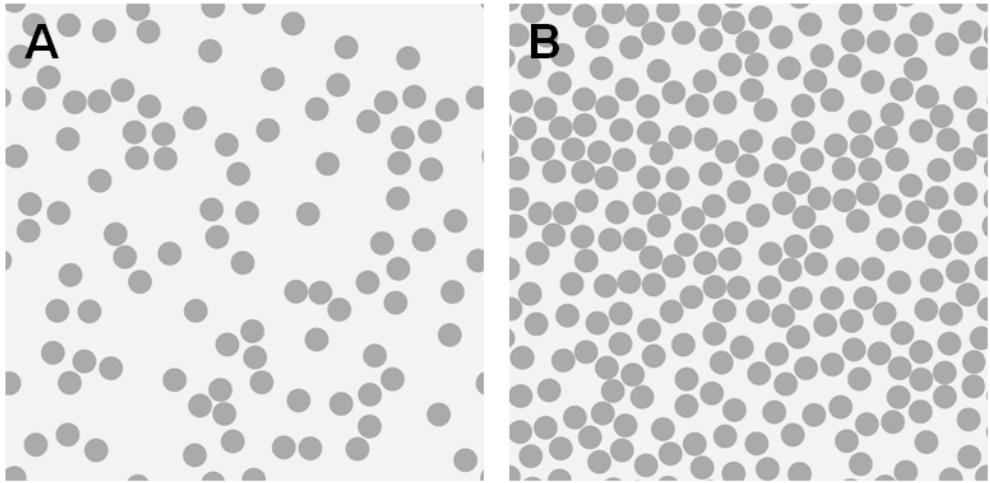

**Figure 1** Snapshots of the binary hard disk fluid. All particles (domains) are identical and non-interacting, so the only parameter is the disk packing fraction $f$ (for these studies, equal to the domain area fraction). Shown are snapshots for packing fraction 0.2 (panel *A*) and 0.5 (panel *B*).

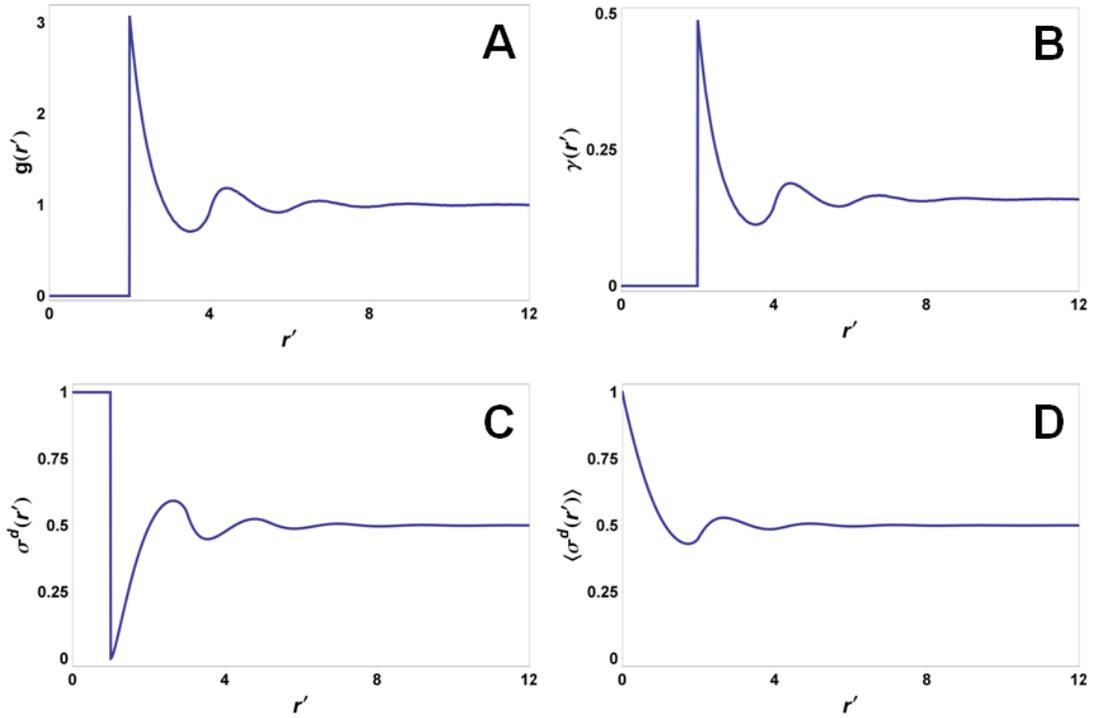

**Figure 2** Domain surface coverage is related to the radial distribution function of the binary hard disk fluid. Shown are intermediate functions from the derivation of the ensemble domain surface coverage. (A) The RDF for disk packing fraction 0.5 shown in Figure 1*B*: $g(r'; 0.5)$. (B) The local number density function corresponding to the RDF in panel *A*: $\gamma(r'; 0.5)$. The local number density function is simply the RDF rescaled by the disk number density of the system, $N = f/\pi$. (C) The domain surface coverage function relative to the center of a reference domain. (D) The ensemble averaged domain surface coverage function, derived by averaging the function in panel *C* over the disk.

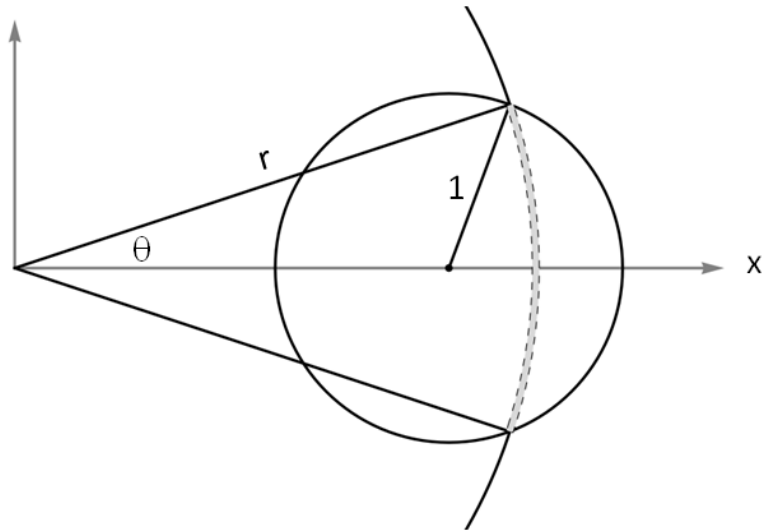

**Figure 3** Geometric considerations for calculating domain surface coverage function $\sigma^d$. The radial distribution function of the system gives the average number of domain *centers* found at a distance $r$ from the origin (taken to be the center of a reference domain), but these are not the only domains contributing to the domain surface coverage at $r$. Shown is a domain centered on the x-axis at $r' < r$, which contributes the shaded area to a thin shell at $r$. It is easily seen that any unit-radius domain whose center lies between $r - 1$ and $r + 1$ will contribute some area to the shell at $r$.

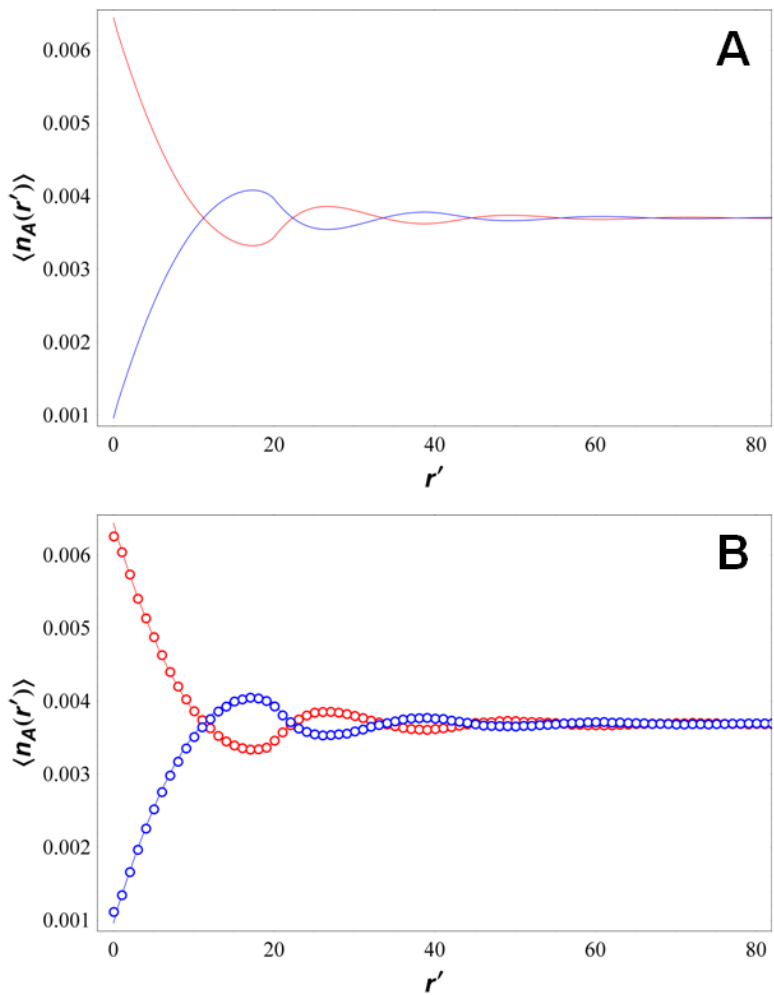

**Figure 4** The average acceptor density seen by donors in a nanodomain system. The two donor pools (those located in domains, and those located in the surrounding phase) see different acceptor densities in the ensemble average. Shown are the acceptor surface densities for domain area fraction 0.4, acceptor mole fraction 0.002, and acceptor partition coefficient of 10 (favoring the domain phase). (A) Acceptor surface density seen by donors in domains (red) and surround (blue) predicted by Equation 20 and 22(B) The same functions shown with MC simulated data, demonstrating the validity of the acceptor density functions.

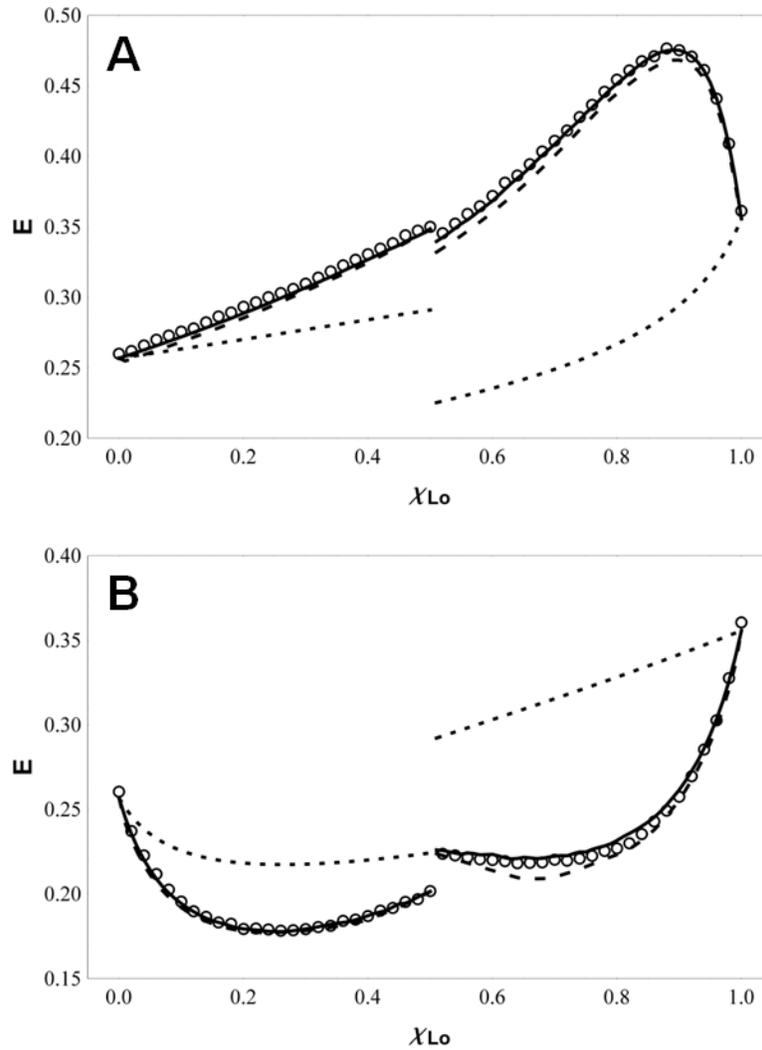

**Figure 5** Comparison of three models for predicting energy transfer efficiency E in bilayers composed of nanoscopic domains. Monte Carlo simulated data (open circles) were generated by constructing random snapshots of domain configurations and randomly placing donors and acceptors, subject to the assumptions of the model. Predictions of the FP-FRET (solid), Towles and Dan (dashed), and Brown (dotted) models. *A*, tieline E profiles for strong donor and acceptor partition into the Ld phase. *B*, tieline E profiles for the case of strong donor partition into Lo, and strong acceptor partition into Ld.